\documentclass[10pt,letterpaper,twocolumn]{article} 

\usepackage{ol}
\usepackage{hyperref}
\usepackage{amsmath}

\begin{document}

\twocolumn[ 

\title{Spontaneous decay rates in active waveguides}

\author{Andr\'es Anibal Rieznik}

\address{Optics and Photonics Research Center, Instituto de F\'{\i}sica Gleb Wataghin, Universidade Estadual de Campinas, cep 13083-970, Campinas, S\~ao Paulo, Brazil, and PADTEC, Rodovia Campinas-Mogi-Mirim (SP 340) Km 118.5, cep 13086-902, Campinas, S\~ao Paulo, Brazil}

\author{Gustavo Rigolin}

\address{Departamento de Raios C\'osmicos e Cronologia, Instituto de F\'{\i}sica Gleb Wataghin, Universidade Estadual de Campinas, C.P. 6165, cep 13084-971, Campinas, S\~ao Paulo, Brazil}


\begin{abstract}
We present a new method to measure the guided, radiated and total decay rates in uniform waveguides. It is also theoretically shown that large modifications of the total decay rate can be achieved in realistic EDFAs and EDWAs with effective mode area radii smaller than $\approx 1 \mu m$.
\end{abstract} 


] 

\noindent It is well-known that the Spontaneous Decay Rates (SDRs) of emitting sources such as atoms or quantum wells can be largely modified in optical micro-cavities. \cite{kleppner} Controlled spontaneous-emission plays a key role in a new generation of micro and nano-optical devices. High-performance micro-cavity lasers, for instance, have been already experimentally demonstrated \cite{yokoyama2,zhang}. SDRs modifications are also expected to impact the performance of optical waveguide amplifiers as the guided mode area radii of these devices become smaller\cite{sondergaarda,rieznik} than $\approx 1 \mu m$. In order to characterize one dimensional nano- and micro-optical devices one key decomposition of the total decay rate ($\tau_{0}^{-1}$) into two components was introduced in Ref. [\citeonline{chu}]. The decay rate was divided into the guided modes ($\tau_{g}^{-1}$) and into the radiated modes ($\tau_{r}^{-1}$), where the total decay rate is $\tau_{0}^{-1} =  \tau_{g}^{-1} + \tau_{r}^{-1}$. In contrast to large devices, where just the total decay rate must be considered for their characterization, the modeling of nano- and micro-optical devices requires the measure of both components of the decay rate. The usual way to determine the SDR of an emitting source embedded in a uniform waveguide consists in the measure of the exponential decay rate of the Amplified Spontaneous Emission (ASE) output power when the pump source is switched off \cite{desurvire}. The SDR is given by the exponential decay coefficient of the ASE output power. 

Two natural questions arise here:(1) How can $\tau_{g}$ and $\tau_{r}$ be measured? (2) What is actually measured when using the classical method to determine the SDR?  In this Letter we answer these two questions. We show that the classical method to measure $\tau_{0}$ in uniform waveguides gives $\tau_{r}$ if used in long length waveguides and actually $\tau_{0}$ in short waveguides (assuming no reflections at the waveguide ends). We also show how these measures are modified in lossy mediums, i. e.,  when a background loss coefficient is incorporated into the rate and propagation equations. Then we show how these ideas are useful on devices of practical interests. Three cases are considered: Erbium Doped Waveguide and Fiber Amplifiers (EDWAs and EDFAs) and Semiconductor Optical Amplifiers (SOAs).

\textit{Theory: decay rate measures and background loss influence}. We employ the analytical solution for the longitudinal $z$ dependence of the rate equations presented in Ref. [\citeonline{rieznik}] to investigate the measure of the SDR in uniform waveguides. The analytical solution presented there is valid only for waveguides in which the excited state population of the emitting source, $N_{2}(z)$, is constant along the fiber. Since the measure of the decay rate is performed when $N_{2} \rightarrow 0$ along $z$, this approximation is valid when measuring $\tau_{0}$. The rate equation is:\cite{rieznik}
\begin{eqnarray}
\frac{\partial \mathcal{N}_{2}(z,t)}{\partial t} & = & - \frac{\mathcal{N}_{2}(z,t)}{\tau_{0}} - \frac{1}{\rho S} \sum_{n = 1}^{M}\left\{ \left[ \left( \alpha_{n} + \gamma_{n}  \right)  \right.\right. \nonumber \\
& & \times \left.\left. \mathcal{N}_{2}(z,t) - \alpha_{n} \right]  P_{n}(z,t) \right\}, \label{rate}
\end{eqnarray}
where $\mathcal{N}_{1} + \mathcal{N}_{2} = 1$ are the normalized population of the upper and lower levels of the emitting source, $\tau_{0}$ is the spontaneous lifetime of the upper level, $\rho$ is the number density of active ions, $S$ is the doped region area, and $\alpha_{n}$ and $\gamma_{n}$ are the absorption and gain constants. The propagation equation is:
\begin{eqnarray}
\frac{\partial {P}_{n}(z,t)}{\partial z} & = & u_{n}\left\{ \left[ \left( \alpha_{n} + \gamma_{n}  \right) \mathcal{N}_{2}(z,t) - \alpha_{n} -\alpha_{loss} \right]  \right. \nonumber \\
& & \times \left. P_{n}(z,t) + 2\gamma_{n}\Delta\nu \mathcal{N}_{2}(z,t) \right\}, \label{propagation}
\end{eqnarray}
where $P_{n}(z,t)$  is the optical power (in photons per unit time) at location $z$ of the $n$th beam with wavelenght centered at $\lambda_{n}$ ($n \leq M$), $u_{n} = 1$ for forward travelling beams and $u_{n} = -1$ for backward travelling beams, $\alpha_{loss}$ is the attenuation coefficient given by the background loss of the fiber glass host, $\Delta \nu$ is the frequency interval between two successive wavelengths considered in the model, and the factor $2$ in the last term stands for two possible polarizations. Solving Eqs.~(\ref{rate}) and (\ref{propagation}) for $\mathcal{N}_{2}(z,t) = \mathcal{N}_{2}(t)$, i. e., $\mathcal{N}_{2}$ constant along $z$, the output power is\cite{rieznik}
\begin{eqnarray}
P^{out}_{n}(t) & = & P^{in}_{n}(t) G_{n}(t) + 2 \mathcal{N}_{n}^{sp} \Delta \nu \left[ G_{n}(t) - 1 \right], \label{output}
\end{eqnarray}
where
\begin{eqnarray}
G_{n}(t) & = & \mathrm{e}^{  (\alpha_{n} + \gamma_{n}) \mathcal{N}_{2}(t)L - (\alpha_{n} - \alpha_{loss}) L }, \label{gain} \\
\mathcal{N}_{n}^{sp} & = &\frac{\gamma_{n} \mathcal{N}_{2}(t)}{(\alpha_{n}+\gamma_{n}) \mathcal{N}_{2}(t) - \alpha_{n} - \alpha_{loss}}. \label{factor} 
\end{eqnarray} 
The rate equation is
\begin{eqnarray}
\frac{\mathrm{d}\mathcal{N}_{2}(t)}{\mathrm{d}t} & = & -\frac{\mathcal{N}_{2}(t)}{\tau_{0}} - \frac{1}{\rho S L}\sum_{n=1}^{M} \left\{ P_{n}^{out}(t) - P_{n}^{in}(t) \right. \nonumber \\
& & \left. - 2 \gamma_{n} \Delta \nu \mathcal{N}_{2}(t) L + \alpha_{loss} H_{n}(t)L\right\}, \label{rate2}
\end{eqnarray}
where
\begin{eqnarray}
H_{n}(t) & = & \frac{P_{n}^{in}(t)}{\ln [G_{n}(t)]}[G_{n}(t) - 1] + 2 \mathcal{N}_{n}^{sp}\Delta \nu \nonumber \\
& & \times \left[ \frac{G_{n}(t) - 1}{\ln [G_{n}(t)]} - 1 \right]. \label{loss}
\end{eqnarray}
Here $P^{out}_{n}(t) = P_{n}(L,t)$ and $P^{in}_{n}(t) = P_{n}(0,t)$ are the output and input power of the $n$th beam, $G_{n}(t)$ is the linear gain, $\mathcal{N}_{n}^{sp}$ is the spontaneous emission factor for the $n$th mode, and $L$ is the doped fiber length.

\textit{Measuring the decay rate}. In the classical method\cite{desurvire} to determine the SDR, the input power is turned off ($P^{in}_{n}(t) = 0$) and  the useful data is collected when the concentration of excited ions is low ($\mathcal{N}_{2} \ll 1$). With these two conditions Eqs.~(\ref{output}) and (\ref{loss}) become:
\begin{eqnarray}
P^{out}_{n}(t) & = & \frac{2 \gamma_{n} \Delta \nu \mathcal{N}_{2}(t)}{\alpha_{n} + \alpha_{loss}} \left[ 1 - \mathrm{e}^{- (\alpha_{n} + \alpha_{loss}) L } \right], \label{output2} \\
H_{n}(t) & = & - \frac{P^{out}_{n}(t)}{L (\alpha_{n} + \alpha_{loss})} + \frac{2 \gamma_{n} \Delta \nu \mathcal{N}_{2}(t)}{\alpha_{n} + \alpha_{loss}}.  \label{loss2} 
\end{eqnarray}
Using Eq.~(\ref{loss2}) we can write Eq.~(\ref{rate2}) as
\begin{eqnarray}
\frac{\mathrm{d}\mathcal{N}_{2}(t)}{\mathrm{d}t} & = & -\frac{\mathcal{N}_{2}(t)}{\tau_{0}} - \sum_{n=1}^{M} \left\{ \frac{P_{n}^{out}(t)}{\rho S L} - \frac{2 \gamma_{n} \Delta \nu \mathcal{N}_{2}(t)}{\rho S}  \nonumber \right. \\
& & \left. - \beta_{n} \left( \frac{P^{out}_{n}(t)}{\rho S L} - \frac{2 \gamma_{n} \Delta \nu \mathcal{N}_{2}(t)}{\rho S} \right) \right\}, \label{rate3}
\end{eqnarray}
where we introduce the effective background loss coefficient for the $n$th mode $\beta_{n} = \alpha_{loss}/(\alpha_{n} + \alpha_{loss})$. Looking at Eq.~(\ref{output2}), which is linear in $\mathcal{N}_{2}(t)$, we see that the right hand side of Eq.~(\ref{rate3}) is also linear in $\mathcal{N}_{2}(t)$. Therefore, it can be rewritten as $\mathrm{d}\mathcal{N}_{2}(t)/\mathrm{d}t = - \mathcal{N}_{2}(t)/\tau_{m}$, where $\tau_{m}$ is what is actually measured by the classical method and not $\tau_{0}$. Since $\tau_{m}$ is quite cumbersome, we do not explictly write it here. But two limiting cases deserve a detailed study.
\textit{Case 1: short length waveguides}. In this case $(\alpha_{n}+\alpha_{loss})L \ll 1$. With this approximation Eq.~(\ref{rate3}) reduces to
\begin{equation}
\frac{\mathrm{d}\mathcal{N}_{2}(t)}{\mathrm{d}t}  = -\frac{\mathcal{N}_{2}(t)}{\tau_{0}}. \label{lpequeno}
\end{equation}
This result shows that only for short length waveguides the classical method\cite{desurvire} furnishes the total SDR of the ion. It is interesting to note that Eq.~(\ref{lpequeno}) is valid whether or not we have background loss ($\alpha_{loss} \neq 0$).
\textit{Case 2: long length waveguides}. Here $(\alpha_{n}+\alpha_{loss})L \gg 1$. Now Eq.~(\ref{rate3}) becomes
\begin{equation}
\frac{\mathrm{d}\mathcal{N}_{2}(t)}{\mathrm{d}t}  = -\frac{\mathcal{N}_{2}(t)}{\tau_{r}} - \sum_{n=1}^{M}\beta_{n}\frac{\mathcal{N}_{2}(t)}{\tau_{g_{n}}}, \label{lgrande}
\end{equation}
where we have used the decomposition\cite{chu} of $\tau_{0}^{-1}$ in guided and radiated modes ($\tau_{0}^{-1} =  \tau_{g}^{-1} + \tau_{r}^{-1}$) and the fact that $\tau_{g}^{-1} = \sum_{i=1}^{M}\tau_{g_{n}}^{-1}$, where $\tau_{g_{n}}^{-1} = 2\gamma_{n}\Delta \nu/\rho S$ is the guided decay rate into the $n$th mode. For sufficiently low background loss $\beta_{n} \approx 0$, which implies that the classical method now furnishes $\tau_{r}^{-1}$.  

The fact that $\tau_{g}^{-1}$ in a given mode can be written as $\tau_{g_{n}}^{-1} = 2\gamma_{n}\Delta \nu /\rho S$ is pointed out here for the first time. It arises naturally from the interpretation given in Ref. [\citeonline{rieznik}] for this term as being the photons captured by the guided modes per unit time. In contrast to Ref. [\citeonline{chu}], in which $\tau_{g}^{-1}$ is given as a function of the dipole moment matrix element between the emitting source excited and ground states, we use here the easily measurable gain constant $\gamma_{n}$. 

Therefore, the total ($\tau_{0}^{-1}$) and the radiated ($\tau_{r}^{-1}$) decay rates in a given uniform waveguide can, in principle, be measured separately employing two different waveguide lengths. $\tau_{g}^{-1}$ can also be determined by simply subtracting the later from the former ($\tau_{g}^{-1} = \tau_{0}^{-1} - \tau_{r}^{-1}$). Anyway, $\tau_{g}^{-1}$ is also easily obtained from the waveguide intrinsic parameters, as discussed in the previous paragraph.

We observe that in Ref. [\citeonline{rieznik}], Section $2.B.3$, the results above  were outlined for $\alpha_{loss} = 0$. But now we explicitly perform the calculations and reinterpret these results in light of the decomposition of the decay rate into guided and radiated modes.

\textit{Simulations}. We end this letter studying waveguide lengths range at which $\tau_{0}^{-1}$ and $\tau_{r}^{-1}$ can be measured. We use three sets of parameters of practical interest. They represent typical EDFAs, EDWAs, and SOAs. It is worth mentioning that the term proportional to $\mathcal{N}_{2}^{2}$, which is usually included in the modelling of EDWAs and SOAs,  can be neglected since $\mathcal{N}_{2} \ll 1$.

\textit{EDFAs and EDWAs.} In usual EDFAs and EDWAs, with optical mode areas larger than $\approx 1 \mu m^{2}$, $\tau_{g}^{-1}$ is negligible and the total lifetime is equal to the radiated lifetime ($\tau_{0}$ $\approx$ $\tau_{r}$ $\approx$ $10 ms$). However, when the optical mode area becomes smaller, $\tau_{g}^{-1}$ starts to have a non-negligible and measurable value. For instance, using the parameters shown in Table \ref{tabela1}, a $1 \mu m$ optical mode radius, and Eq.~(\ref{rate3}), we determine the measured lifetime as a function of the waveguide length for typical EDFAs and EDWAs (assuming perfect detection). Without loss of generality, we use an effective gain and absorption constant along the total transition bandwidth of $\approx 15 Thz$ centered at $1.55 \mu m$. 
\begin{table}[!ht]
\vspace{-.5cm}
\centering
\caption{\footnotesize{Parameters used in the simulations.}}
\vspace{-.25cm}
\begin{tabular}{ccccc} \\ \hline \hline
Parameter & $\gamma_{eff}$ & $\alpha_{eff}$ & $\tau_{r}$ & $\rho$ \\ 
 & ($m^{-1}$) & ($m^{-1}$) & ($ms$) & ($m^{-3}$) \\ \hline
EDFA\cite{rieznik} &  $0.2$ & $0.2$ & $10$ & $1.0 \times 10^{24}$ \\
EDWA\cite{snoeks}  & $20$ & $49$ & $22$ & $1.4 \times 10^{26}$ \\ \hline \hline
\end{tabular}
\label{tabela1}
\end{table}
\begin{figure}[!htb]
\vspace{-.25cm}
\centerline{\includegraphics[width=8cm,height=6cm]{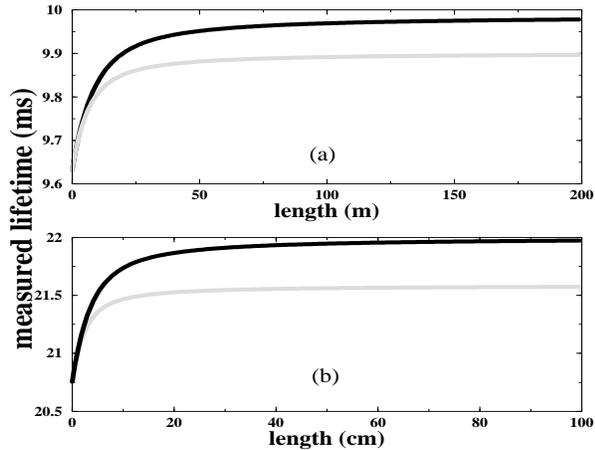}}
\vspace{-.35cm}
\caption{\footnotesize{Simulations of the measured lifetimes as a function of the waveguide length for typical EDFAs (a) and EDWAs (b). The black lines are for non-lossy waveguides and the light grey ones for $\alpha_{loss}=0.3 dB/m$ (EDFAs)\cite{georges} and $1 dB/cm$ (EDWAs)\cite{snoeks}, typical of fluorozirconate EDFAs and silica based EDWAs, respectively.}}
\label{fig1}
\vspace{-.3cm}
\end{figure}
We observe in Fig.~\ref{fig1} that a $1 \mu m$ optical mode radius is small enough to cause a variation of $\approx 5 \%$ between the total and the radiated lifetimes, measured at small and large waveguide lengths, respectively. Although small, an optical mode radius of $\approx 1 \mu m$ is already commercially available in Photonic Crystal Fibers. Moreover, it has been shown that waveguides with high index-contrast (and, consequently, small mode areas) have several advantages\cite{saini}, which envisages the future construction of very-small mode area devices. To study how the total and radiated decay rates can be measured in such devices, we perform simulations assuming the values shown in Table \ref{tabela1} for an EDWA, but an optical mode area of $0.02 \mu m^{2}$, which was obtained, for instance, in Ref. [\citeonline{zhang}]. The results are shown in Fig.~\ref{fig2}. Of course, such highly-confined EDWA would use materials which will not necessarily have the parameters shown in Table \ref{tabela1}. But the graphic in Fig.~\ref{fig2} is illustrative of the effects that would always occur at smaller mode areas: the difference between the total and the radiated lifetime increases (as a consequence of larger $\tau_{g}^{-1}$), the difference between the decay rates in lossy and  non-lossy waveguides increases, and, at last, the waveguide lengths to obtain the radiated lifetime with a given accuracy also increases. 
\begin{figure}[!htb]
\centerline{\includegraphics[width=8cm,height=3cm]{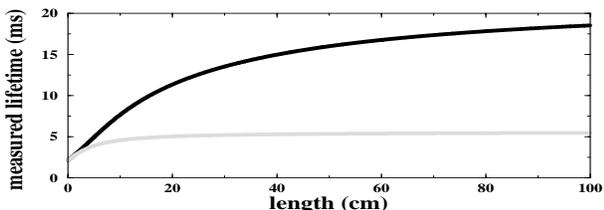}}
\vspace{-.5cm}
\caption{\footnotesize{Theoretical lifetimes as a function of the waveguide lenght for an EDWA with $0.02 \mu m^{2}$ optical mode area. Other parameters are given in Tab.~\ref{tabela1}.}}
\label{fig2}
\vspace{-.5cm}
\end{figure}

\textit{SOAs.} In SOAs, the condition $\mathcal{N}_{2}$ $+$ $\mathcal{N}_{1}$ $=$ $1$ does not hold, but it is easy to show that the method here presented to measure the radiated and total lifetime also works. We observe that $\gamma_{n}$ can be written as $\sigma_{n}^{e}\Gamma \rho$ $/S$, where $\sigma_{n}^{e}$ is the emission-crosssection at wavelength $\lambda_{n}$ and $\Gamma$ is the overlap factor between the optical mode and the doped region area\cite{giles}. Then, $\tau_{g}^{-1}$ is given by $\tau_{g}^{-1}= 2 \sigma_{n}^{e} \Gamma /S$. Using $\sigma_{n}^{e}$ $\Gamma$ $=$ $2.0 \times 10^{-16}$ $cm^{2}$ for a typical SOA\cite{agrawal}, we found that $\tau_{g}^{-1}$ becomes comparable to $\tau_{r}^{-1}$ of $\approx (200 ps)^{-1}$ only at optical mode radius smaller than $\approx 1 nm$, far from the possibilities of present technologies. 

G. Rigolin thanks FAPESP for funding this research.

\end{document}